\documentclass[twocolumn,amsmath,amssymb,pr,superscriptaddress]{revtex4}
\usepackage{graphicx}
\usepackage{color}
\usepackage{hyperref}
\usepackage[abs]{overpic}
\usepackage{pict2e}
\usepackage{float}

\begin{document}
\title{Phase Diagram of $\alpha$-RuCl$_3$ in an in-plane Magnetic Field}

\author{J. A. Sears}
\affiliation{Department of
Physics, University of Toronto, 60 St.~George St., Toronto, Ontario,
M5S 1A7, Canada}
\author{Y. Zhao}
\author{Z. Xu}
\affiliation{NIST Center for Neutron Research, National Institute of Standards and Technology, Gaithersburg, Maryland, 20899, USA}
\affiliation{Department of Materials Science and Engineering, University of Maryland, College Park, Maryland 20742, USA}
\author{J. W. Lynn}
\affiliation{NIST Center for Neutron Research, National Institute of Standards and Technology, Gaithersburg, Maryland, 20899, USA}
\author{Young-June~Kim}
\email{yjkim@physics.utoronto.ca} \affiliation{Department of
Physics, University of Toronto, 60 St.~George St., Toronto, Ontario,
M5S 1A7, Canada}

\date{\today}

\begin{abstract}
The low-temperature magnetic phases in the layered honeycomb lattice material $\alpha$-RuCl$_3$ have been studied as a function of in-plane magnetic field. In zero field this material orders magnetically below 7 K with so-called zigzag order within the honeycomb planes. Neutron diffraction data show that a relatively small applied field of 2 T is sufficient to suppress the population of the magnetic domain in which the zigzag chains run along the field direction. We found that the intensity of the magnetic peaks due to zigzag order is continuously suppressed with increasing field until their disappearance at $\mu_o$H$_c$=8 T. At still higher fields (above 8 T) the zigzag order is destroyed, while bulk magnetization and heat capacity measurements suggest that the material enters a state with gapped magnetic excitations. We discuss the magnetic phase diagram obtained in our study in the context of a quantum phase transition.
\end{abstract}

%\pacs{75.10.Jm, 75.25.-j, 75.40.Cx}

\maketitle

The transition metal halide $\alpha$-RuCl$_3$ has a crystal structure made up of stacked honeycomb layers of edge-sharing RuCl$_6$ octahedra. Plumb et al. \cite{plumb14} found that spin orbit coupling in this material is substantial, leading to a j$_{eff}$=$\frac{1}{2}$ state description of the Ru$^{3+}$ valence electrons. Since this material is built up with edge-sharing RuCl$_6$ octahedra, its spin Hamiltonian is believed to include a significant bond-dependent Kitaev interaction \cite{kitaev2006,jackeli2009}, making $\alpha$-RuCl$_3$ a material of great interest in the ongoing search for a Kitaev spin liquid ground state \cite{sandilands2016,koitzsch2016,lang2016,sinn2016,weber2016,zhou2016,hirobe2016,lampenkelley2016,park2016,hsdo2017,hskim15,nasu2016,chaloupka2016,sizyuk2016,yadav16,catuneanu2017,hou2016,chern2016,gohlke2017,trebst2017}. Although $\alpha$-RuCl$_3$ orders magnetically at low temperature with zigzag magnetic order \cite{sears15, banerjee16, johnson15, cao16}, this material has shown some signatures of spin-liquid physics, such as a broad continuum of magnetic excitations identified in both Raman scattering \cite{sandilands15} and inelastic neutron scattering measurements \cite{banerjee16, banerjee16a}.

When a magnetic field is applied within the honeycomb plane, previous bulk measurements have reported that $\alpha$-RuCl$_3$ undergoes a number of transitions \cite{kubota15, majumder15, johnson15, baek2017}, including low field transitions resembling spin-flop transitions occurring at 1~T and 6~T, followed by the apparent loss of zigzag magnetic order at 8~T. In contrast, when a magnetic field is applied perpendicular to the honeycomb planes the zigzag magnetic order appears to be robust up to fields of 14~T \cite{majumder15}. The high field phase above the loss of zigzag magnetic order has been the subject of particular interest recently \cite{johnson15,kubota15,majumder15,baek2017,leahy2016}. It has been proposed that this phase may be a simple polarized paramagnetic state \cite{johnson15}, however this does not account for the lack of saturation in the magnetization \cite{kubota15}. The high field phase has also been characterized by NMR measurements \cite{baek2017} which show that the magnetic excitations develop an energy gap. The gap size was similar for the two field directions measured, a result difficult to reconcile with the physics of a polarized paramagnetic state. This finding of gapped excitations in the high field phase is in contrast to recent thermal conductivity measurements \cite{leahy2016}, which suggested the presence of gapless excitations in the high field phase.

In this paper, we have characterized these finite field transitions using magnetic neutron diffraction, and bulk heat capacity and magnetization measurements on the same samples. Neutron diffraction measurements show that at low field (2~T) the diffracted intensity due to one of the zigzag domains disappears, suggesting that redistribution in domain population occurs in this rather low field range. We found that the zigzag magnetic order temperature $T_c$ is continuously suppressed with applied field, and eventually disappears above the critical in-plane field of ${\mu_o H}_{c}=8$~T.\footnote{We confine our discussion only to magnetic field applied within the honeycomb plane. Justification for this is purely empirical. Large anisotropy in magnetic susceptibility measurements and high-field magnetization indicates that easy axis is within the honeycomb plane. However, a recent neutron diffraction study reported that the ordered moment has components perpendicular to the plane. We are only concerned with magnetic behavior projected to the honeycomb plane in this paper.}
The high field phase above the critical field is characterized by a magnetic excitation gap, $\Delta$, which can be extracted from the specific heat data. The energy scales both below ($T_c$) and above ($\Delta$) the critical field exhibit power law scaling as a function of in-plane magnetic field, indicating the presence of a quantum critical point. We note that the quantum phase transition due to transverse-field in the Ising model provides a reasonable phenomenological description of the observed phase diagram.

\begin{figure*}
  \centering
  \begin{overpic}[scale=0.5,unit=1mm]{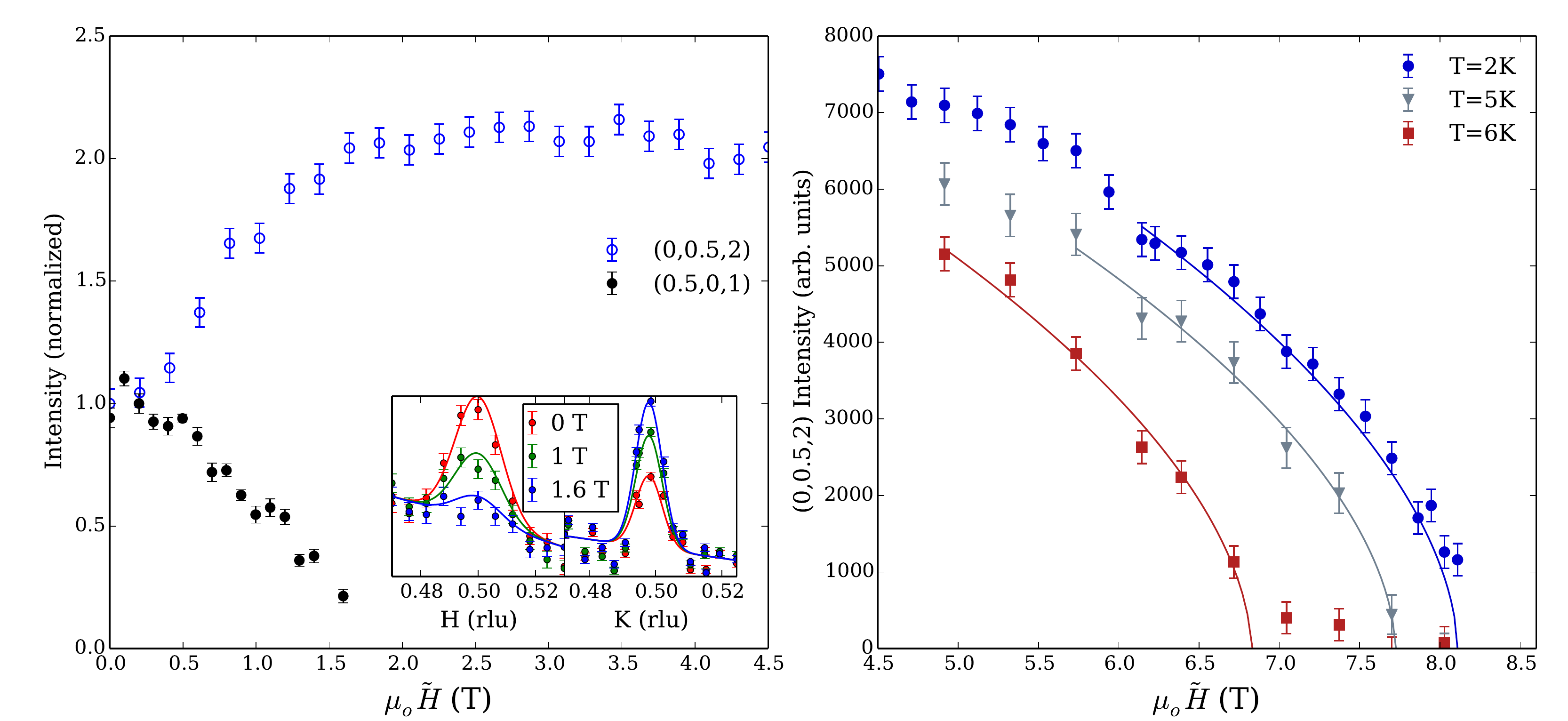}
    \put(20,68){(a)}
    \put(105,68){(b)}
  \end{overpic}
  \caption{(a) Low field magnetic peak intensity at 2 K as a function of the in-plane component of magnetic field ($\mu_o \tilde{H}$). The intensity is normalized to the value at zero field. The inset shows individual scans of (0.5,0,1) and (0,0.5,2) Bragg peaks at 0, 1, and 1.6 T (in-plane field) and 2 K. (b) High field intensity of the (0,0.5,2) peak at 2 K, 5 K, and 6 K. Solid lines are fits with $\sim ({H}-{H}_c)^{2 \beta^{*}}$ to extract the critical field. Same critical exponent $\beta^{*} = 0.28$ was used for all three curves. Error bars where indicated represent one standard deviation.}
\label{fig:neutron}
\end{figure*}

\begin{figure}
  \begin{minipage}[b]{0.14\textwidth}
      \centering
      \hspace*{0.49cm}
      \includegraphics[width=0.9\textwidth, angle=90]{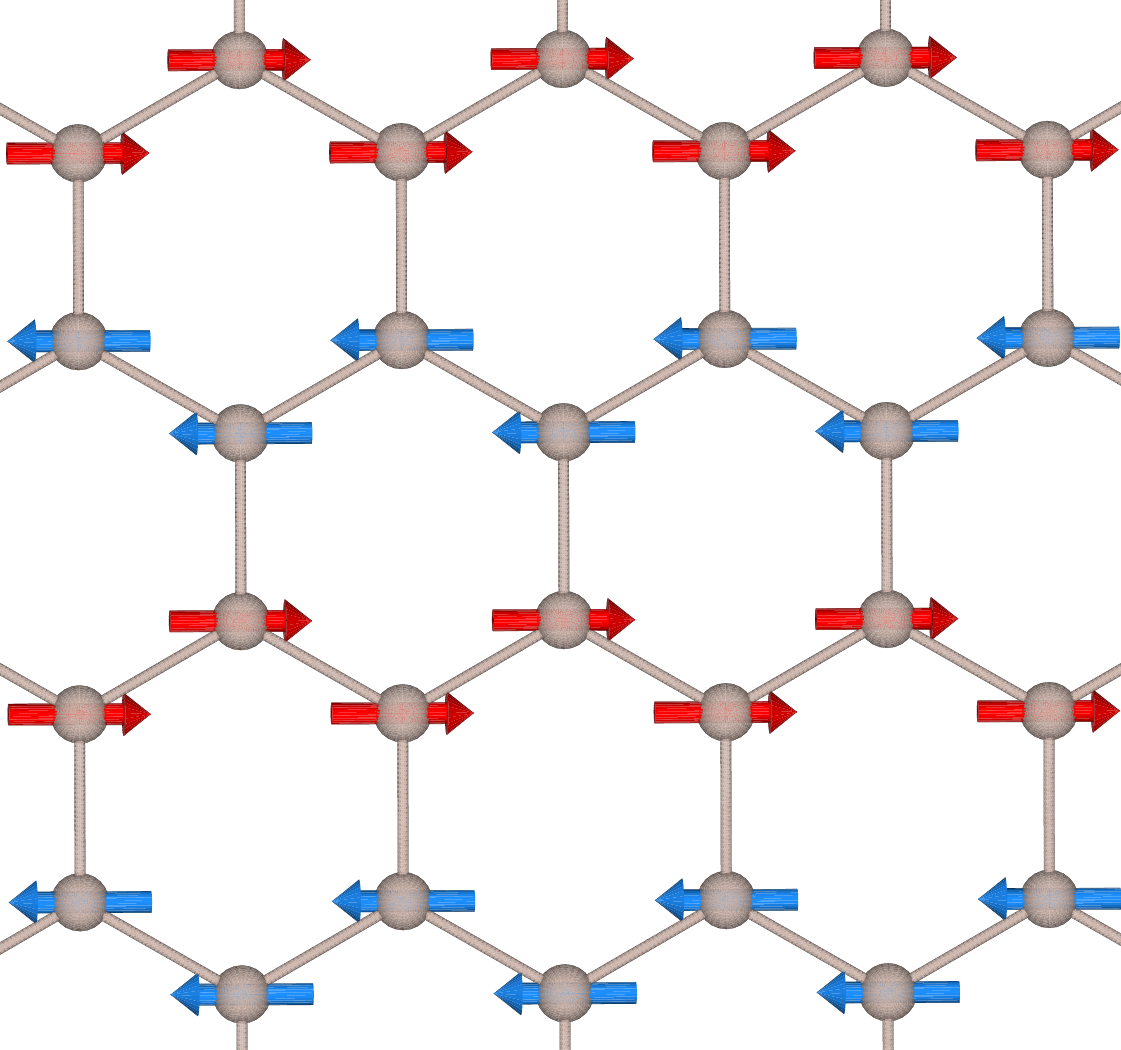}
  \end{minipage}
  \begin{minipage}[b]{0.14\textwidth}
      \centering
      \hspace*{0.5cm}
      \includegraphics[width=0.9\textwidth, angle=90]{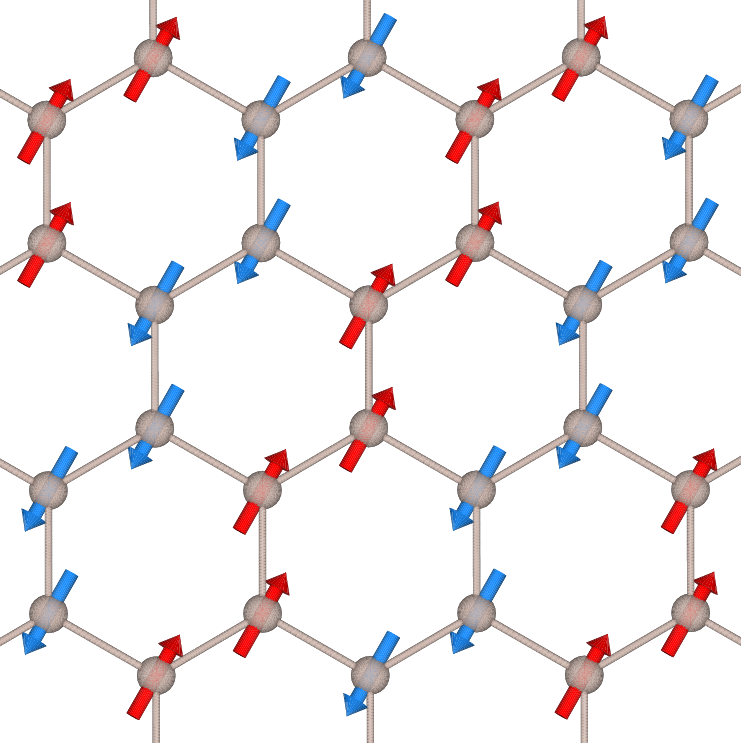}
  \end{minipage}
  \begin{minipage}[b]{0.14\textwidth}
      \centering
      \hspace*{0.6cm}
      \includegraphics[width=0.9\textwidth, angle=90]{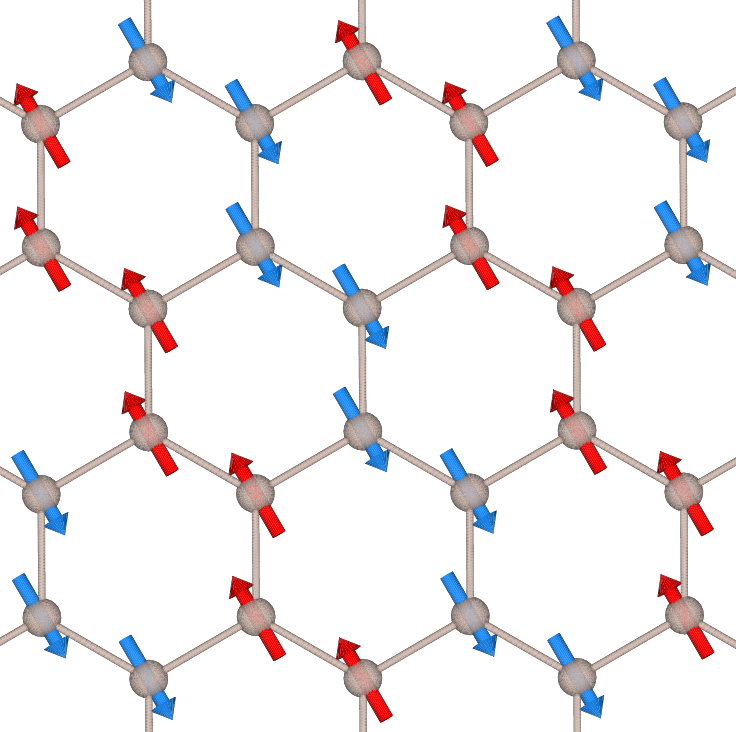}
  \end{minipage}
  \begin{minipage}[b]{0.5\textwidth}
     \centering
      \begin{overpic}[width=1.0\textwidth]{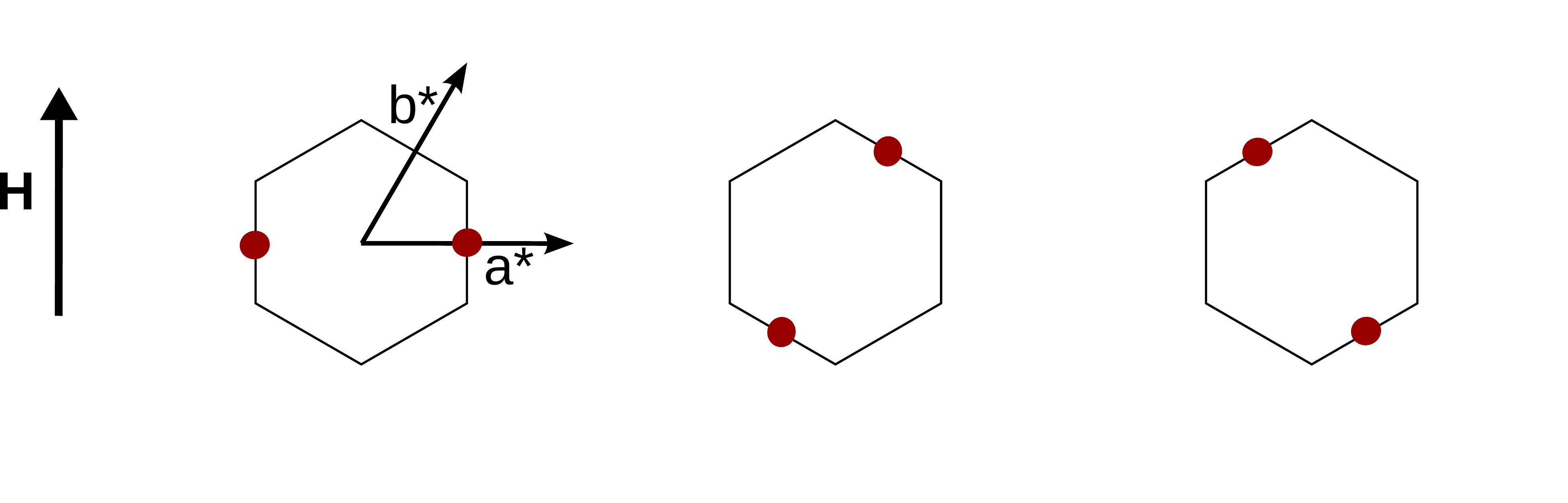}
        \put(39,-3){Domain 1}
        \put(117,-3){Domain 2}
        \put(193,-3){Domain 3}
      \end{overpic}
  \end{minipage}
\caption{Magnetic structures and Bragg peak positions in the first Brillouin zone for each of the three possible zigzag magnetic domains. In a vertical magnetic field the intensities due to Domain 1 disappeared and intensities due to Domain 2 increased. Note that the moments are shown pointing along the zigzag direction for illustrative purposes only. Drawings of magnetic structure were done in VESTA 3 \cite{momma2011}.}
\label{fig:reci}
\end{figure}

Single crystals of $\alpha$-RuCl$_3$ were grown from commercial RuCl$_3$ powder (Sigma-Aldrich, Ru content 45-55\%) by vacuum sublimation in sealed quartz tubes. This resulted in flat, plate-like crystals with typical dimensions 1-2 mm$^2$ and mass 1-5 mg. The crystallographic c direction (hexagonal notation) was found to be perpendicular to the large surface of the crystal. Throughout this paper we will use the hexagonal crystallographic notation with $a=5.96$~\AA\ and $c=17.2$~\AA, in which the a-b plane coincides with the honeycomb layers. The crystals have well defined facets at 120$^{\circ}$ angles, and it was found that the facets coincide with the hexagonal (1,1,0) type directions.

Neutron diffraction measurements were carried out using the BT-7 triple axis spectrometer at the NIST Center for Neutron Research (NCNR) \cite{lynn2012}. The neutron diffraction data were collected using a crystal array of 60 crystals, with a mass of 100 mg. The incident neutron energy was 14.7 meV, and measurements were conducted in the (H0L) plane as well as the plane containing the (0, 0.5, 2) and (-0.5, 0.5, 2) magnetic Bragg peaks. In both cases magnetic fields up to 15~T were applied perpendicular to the scattering plane using either a 10~T or a 15~T vertical field superconducting magnet. In order to gain access to the (0,0.5,2) magnetic peak it was necessary to rotate the sample such that the angle between the magnetic field and and the honeycomb plane was approximately 35$^{\circ}$. In this case, we quote the in-plane component of the field, $\tilde{H}$, rather than the total field applied. For all the other measurements, magnetic field was applied within the honeycomb plane.

Magnetization and heat capacity was measured as a function of temperature using a Physical Property Measurement System (PPMS) with fields up to 14~T. The magnetization measurements were conducted on a collection of six crystals mounted with the field applied along the in-plane (-1,2,0) direction. The heat capacity measurements were done with a single crystal mounted vertically on an aluminum oxide mount in the same orientation as that used for the magnetization measurements. The phonon contribution to heat capacity was subtracted using the non-magnetic isostructural $\alpha$-IrCl$_3$ \cite{ircl3}.

\begin{figure}
  \centering
  \begin{overpic}[width=0.5\textwidth]{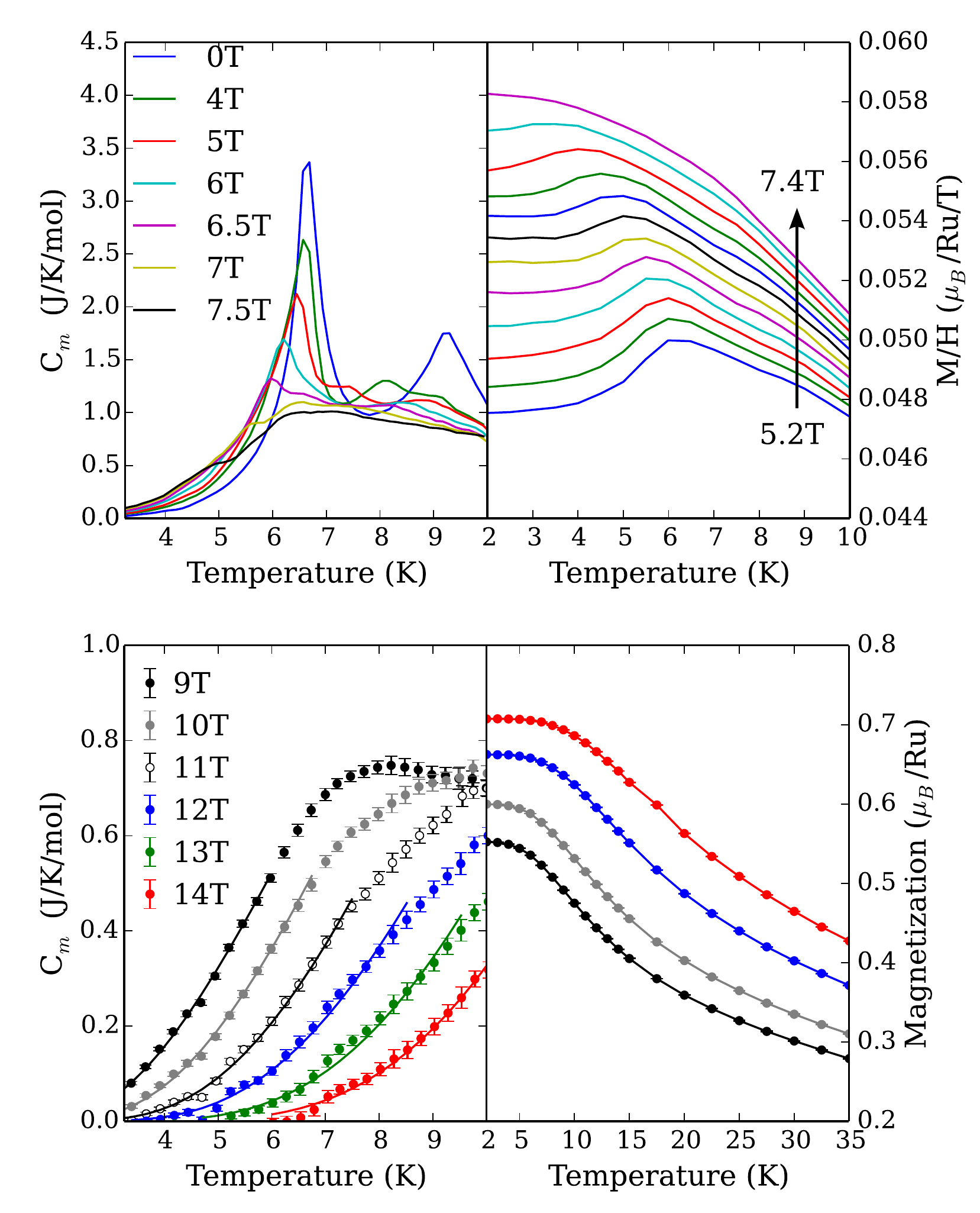}
    \put(100,288){(a)}
    \put(200,288){(b)}
    \put(100,134){(c)}
    \put(200,134){(d)}
  \end{overpic}
  \caption{Magnetic heat capacity as a function of temperature and in-plane magnetic field (a) below the 8~T transition and (c) above 8~T. The phonon contribution was removed by subtracting the heat capacity of isostructural $\alpha$-IrCl$_3$. The solid lines in (c) are fits to an exponential expression for a gapped system ($ A e^{-\Delta /T}$). (b) Magnetization divided by magnetic field as a function of temperature for magnetic fields ranging from 5.2 to 7.4~T in steps of 0.2~T.  (d) Magnetization as a function of temperature for magnetic fields above the 8~T transition.}
\label{fig:cp}
\end{figure}
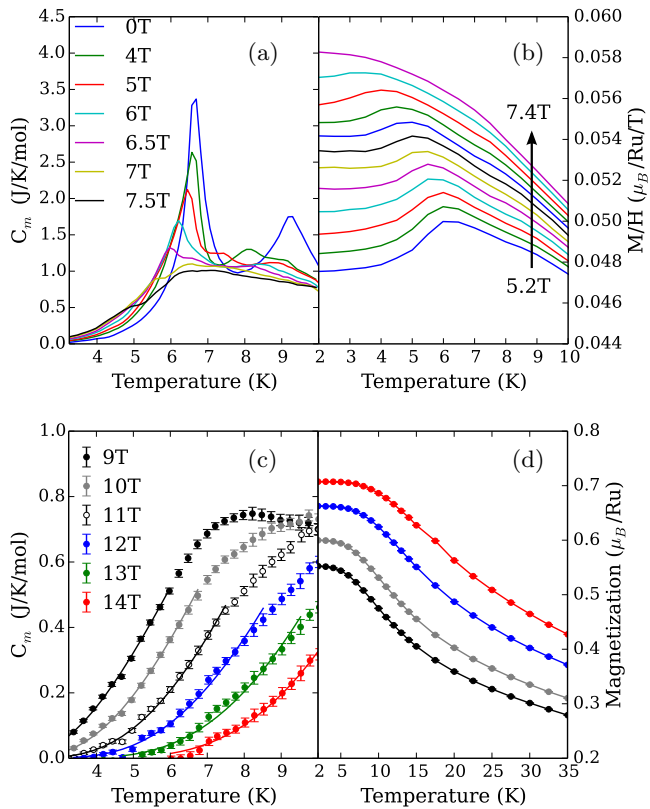

We have investigated the magnetic transitions directly by measuring the magnetic Bragg peak intensity as a function of field. When the magnetic field was applied perpendicular to the H0L plane, all the magnetic peaks in this plane -- $(\pm0.5,0,l)$ with $l=1,2,4$ -- decreased in intensity and disappeared at the relatively low magnetic field of 2~T. The sample was then rotated to gain access to the (0,0.5,2) magnetic peak, which was found to increase in intensity over this field range, as shown in Fig.~\ref{fig:neutron}(a). The critical field for this transition is in rough correspondence with the low field transition observed in bulk measurements and previously interpreted as a spin-flop type transition, which traditionally refers to a re-orientation of spins perpendicular to the applied field in an antiferromagnet \cite{kubota15}. In our experiment, the direction perpendicular to the applied field corresponds to the hexagonal (1,0,0) direction, which is not one of the easy-axes. We also note that a spin-flop transition still preserves the magnetic ordering wave vector, even though magnetic Bragg peak intensities will be modified. Therefore, spin-flop transition is not compatible with our observation of the disappearance of {\em all} magnetic Bragg peaks in the H0L plane. This unexpected finding can be explained as a result of a change in magnetic domain population. Zigzag magnetic order can be described as ferromagnetic zigzag chains, running along the so-called zigzag direction of a honeycomb lattice, coupled antiferromagnetically. Due to the 3-fold symmetry of the lattice, zigzag magnetic order may occur in one of three possible directions, resulting in three magnetic domains that contribute to diffraction intensity in different regions of reciprocal space as shown in Fig.~\ref{fig:reci}. The disappearance of the peaks in the H0L plane is well explained by the disappearance of domain 1 as shown in Fig.~\ref{fig:reci}. The increase in intensity for the (0,0.5,2) magnetic peak belonging to domain 2 is expected for a redistribution of domain population from domain 1 into domains 2 and 3. We confirmed that the domain 3 population increases with field as well (not shown). We note that this ``domain-reorientation" occurs gradually with field, and reaches equilibrium above about 2~T. The observed {\em gradual} field-dependence is also consistent with this change coming from domain population change as spin-flop transitions tend to be first order when the field is parallel to the spin direction. Above this ``domain-reorientation'' transition, the magnetic Bragg peak shows little change in intensity up to 6~T. Above 6~T the intensity begins to decrease and disappears entirely above $\mu_o {H}_c \approx8$~T, directly confirming that zigzag magnetic order disappears above a critical in-plane field of approximately 8~T. This transition is continuous as a function of magnetic field. The zigzag order parameter, $\sqrt{I}$, where $I$ is the intensity of the (0,0.5,2) peak, exhibits power law behavior $\sqrt{I} \sim ({H}-{H}_c)^{\beta^{*}}$ with $\beta^*=0.28 \pm 0.05$. This power law behavior seems to hold for higher temperature data as well, although the critical field H$_c$ shifts to lower field with increasing temperature.

Heat capacity and magnetization data collected at zero magnetic field both show signatures of the zigzag magnetic ordering at low temperature. The heat capacity at zero magnetic field shows a sharp feature at 6.5~K and a second, smaller feature around 9~K [Fig.~\ref{fig:cp}(a)]. The magnetic Bragg peaks observed by neutron diffraction in our samples show an ordering temperature of about 7-8~K \cite{sears15}, so we attribute the lower temperature feature to this zigzag ordering. The nature of the 9~K feature seen in our samples is not known, but Cao et al. reported that stacking disorder in $\alpha$-RuCl$_3$ can increase the ordering temperature to approximately 14~K \cite{banerjee16, cao16} and it is plausible to suppose that the 9 K transition observed in our sample arises from a grain with a different stacking order. 

As field is increased, the sharp feature in the heat capacity decreases in size before shifting to lower temperature and becoming difficult to resolve as shown in Fig.~\ref{fig:cp}(a). Magnetization data at low field show a sharp drop upon decreasing temperature below 7~K as the crystal enters the ordered phase. Figure~\ref{fig:cp}(b) shows that this drop becomes smaller in size and eventually disappears at high field once the zigzag magnetic ordering has disappeared. In the high field phase the heat capacity no longer shows any sharp feature, but instead shows a broad feature that increases in temperature with increasing magnetic field [Fig.~\ref{fig:cp}(c)]. The low temperature heat capacity data were fit using an expression for activated behavior ($ A e^{-\Delta /T}$) to extract the magnetic excitation gap $\Delta$. The magnetization in the high field phase shown in Fig.~\ref{fig:cp}(d) increases gradually with decreasing temperature, reaching field-dependent saturation values at low temperature.

\begin{figure}
  \centering
  \includegraphics[width=0.4\textwidth,angle=-90]{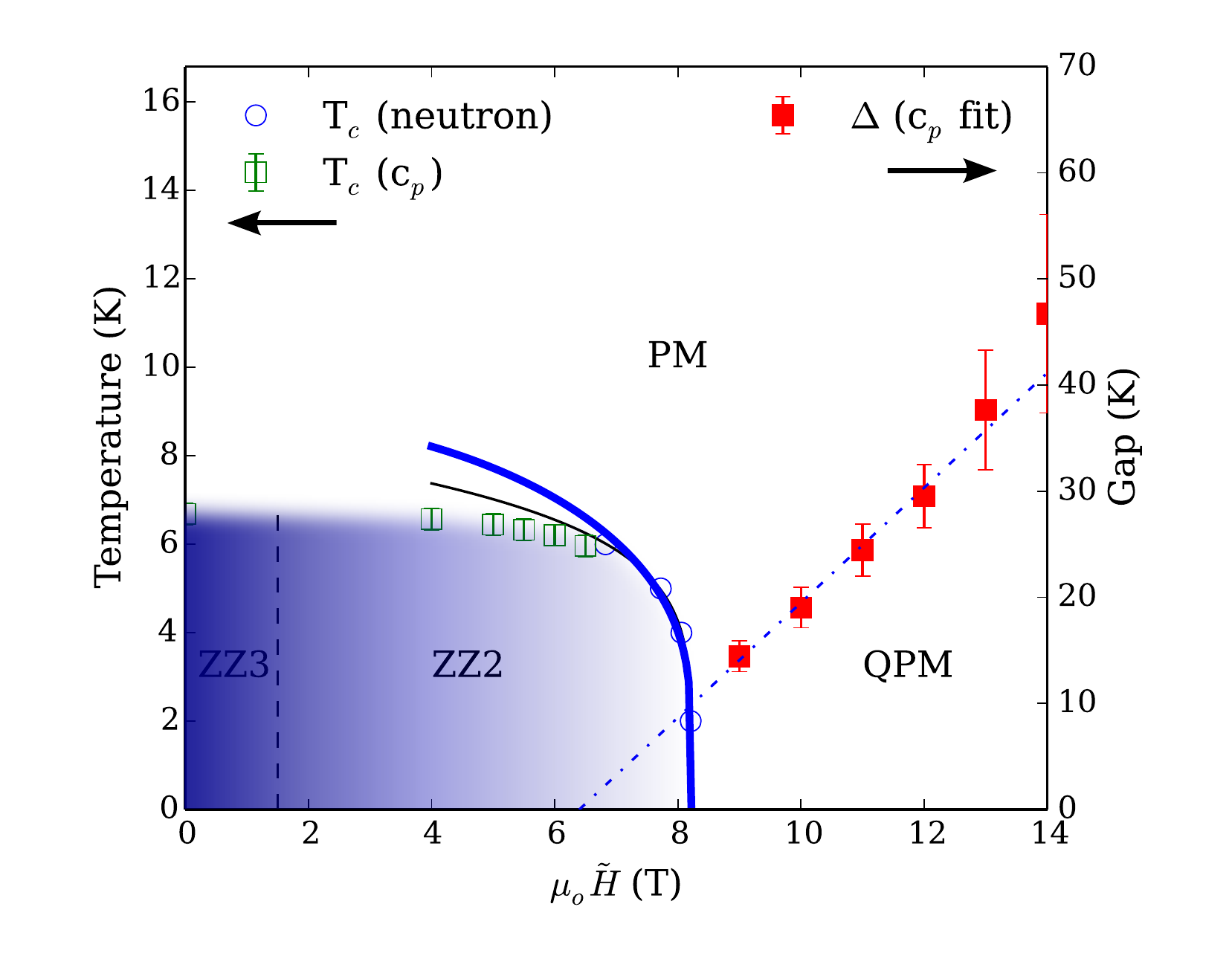}
  \caption{In-plane field -- temperature phase diagram. ZZ3: zigzag magnetic order with three equal domain populations; ZZ2: zigzag magnetic order with redistributed (two) domain population; QPM: quantum disordered phase with gapped magnetic excitations; PM: paramagnetic phase. The phase boundary between ZZ2 and PM is the transition temperature $T_c$ obtained from heat capacity and neutron measurements. The thick solid line is from the transverse field Ising model, and the thin solid line is fit with a power law as described in the text. The value of $\Delta$ found from the heat capacity data is also shown (right-hand axis) and the dashed line is a linear fit to the gap size $\Delta$. }
  \label{fig:phase}
\end{figure}

The experimental results are summarized in Fig.~\ref{fig:phase}, which combines neutron and bulk measurements to determine the phase diagram. The low field transition was found to be a change in magnetic domain population, separating phases made up of 3 and 2 magnetic domains (phases ZZ3 and ZZ2 respectively). The loss of magnetic order above the high field transition was also confirmed, although the nature of the high field phase remains to be clarified. The magnetic excitation gap in the high field phase was characterized by fitting low temperature heat capacity data. The gap size scales with magnetic field, going to zero at finite field rather than at zero field as would be expected for a simple polarized paramagnetic state. This finding is consistent with the NMR measurements reported previously \cite{baek2017}, but contrasts with the results of thermal conductivity measurements which suggested the presence of gapless excitations \cite{leahy2016}.

The observation of  vanishing energy scales towards a critical field in both high and low field regimes is strongly suggestive of quantum critical behavior. Although detailed analysis of the spin Hamiltonian of $\alpha$-RuCl$_3$ is beyond the scope of this paper, the phase diagram could be understood heuristically by comparing our results with one of the simplest models that goes through a quantum phase transition: the transverse-field Ising model (TFIM). There is also physical motivation for our choice of transverse field Ising model. $\alpha$-RuCl$_3$ does show a large uniaxial anisotropy and the magnetic field in our experimental setup has a large component transverse to the easy axis. This is a result of the domain-reorientation transition which favors domains in which the zigzag chain directions are perpendicular to the field direction. The moment direction has been found to point along the zigzag direction (neglecting a small out-of-plane component) \cite{cao16}, resulting in a phase with magnetic field nearly perpendicular to the moment directions.

In Fig.~\ref{fig:phase}, we compare the phase boundary with the TFIM mean field result and find that they are in good agreement in the region close to $H_c$. We could also fit $T_c(H)$ using a power law with $T_c(H) \sim (H_c-H)^{0.18}$ as shown in the figure. Above the critical field, the gap follows a power law scaling $\Delta \sim (H-H_c^*)^{z \nu}$ with $z \nu \approx 1$.  Note that the critical field value extrapolated from this scaling $H_c^* \approx 6.5$~T is slightly different from the critical field $\mu_o H_c \approx 8$~T.  This discrepancy may be due to the complex nature of the Hamiltonian of the real material, or indicates the necessity of another parameter that needs to be tuned to reach the quantum critical point that exists away from the $T-H$ plane. We note that the critical exponent relation $z \nu=1$ is consistent with the $d=2$ Ising model \cite{sachdev2001}. In addition, in Fig.~\ref{fig:neutron}(b), the magnetic order parameter could be fitted well using the critical exponent $\beta^*=0.28$, which is close to the theoretical value of 0.32 \cite{pfeuty1976}. Finally, the low-temperature saturation behavior observed in Fig.~\ref{fig:cp}(d) is naturally explained by the temperature dependence of the transverse magnetization in the TFIM.

In conclusion, we have determined the high field phase diagram for $\alpha$-RuCl$_3$ using neutron diffraction, magnetization, and heat capacity measurements. We have confirmed the loss of zigzag order in the high field phase and found that the material enters into a phase with gapped magnetic excitations. The experimentally determined energy scales represented by the magnetic ordering temperature for fields below the critical field and the energy gap above the critical field, both show power law scaling behavior and vanish towards the critical field, indicating a field-driven quantum phase transition. We found that the phase diagram and the critical behavior is qualitatively similar to that expected for the transverse field Ising model.

Research at the University of Toronto was supported by Natural Science and Engineering Research Council of Canada through Discovery Grant and Collaborative Research and Training Experience (CREATE) program. The identification of any commercial product or trade name does not imply endorsement or recommendation by the National Institute of Standards and Technology.

\bibliography{rucl3_exp,rucl3_theory,rucl3_exp_pre,rucl3_theory_pre}

\end{document}